\documentclass[pdflatex,sn-mathphys-num]{sn-jnl}

\usepackage{graphicx}%
\usepackage{multirow}%
\usepackage{amsmath,amssymb,amsfonts}%
\usepackage{amsthm}%
\usepackage{mathrsfs}%
\usepackage[title]{appendix}%
\usepackage{xcolor}%
\usepackage{textcomp}%
\usepackage{manyfoot}%
\usepackage{booktabs}%
\usepackage{algorithm}%
\usepackage{algorithmicx}%
\usepackage{algpseudocode}%
\usepackage{listings}%
\usepackage{tabularray}
\usepackage{tabularx}

\theoremstyle{thmstyleone}%
\theoremstyle{thmstyletwo}%

\theoremstyle{thmstylethree}%

\begin{document}

\title{Multisensory extended reality applications offer benefits for volumetric biomedical image analysis in research and medicine\protect\footnote{This version of the article has been accepted for publication, after peer review but is not the Version of Record and does not reflect post-acceptance improvements, or any corrections. The Version of Record is available online at: \url{http://dx.doi.org/10.1007/s10278-024-01094-x}}}

\author*[1,2]{\fnm{Kathrin} \sur{Krieger}}\email{kathrin.krieger@isas.de} 
\author[3,4]{\fnm{Jan} \sur{Egger}}\email{jan.egger@uk-essen.de} 
\author[3]{\fnm{Jens} \sur{Kleesiek}}\email{jens.kleesiek@uk-essen.de} 
\author[1,5]{\fnm{Matthias} \sur{Gunzer}}\email{matthias.gunzer@isas.de} 
\author[1]{\fnm{Jianxu} \sur{Chen}}\email{jianxu.chen@isas.de} 

\affil*[1]{\orgdiv{Biospectroscopy}, \orgname{Leibniz-Institut for Analytical Science-ISAS-e.V.}, \orgaddress{\street{Bunsen-Kirchhoff-Str. 11}, \city{Dortmund}, \postcode{44139}, \state{NRW}, \country{Germany}}}
\affil[2]{\orgdiv{Neuroinformatics Group, Faculity of Technology}, \orgname{Bielefeld University}, \orgaddress{\street{Inspiration 1}, \city{Bielefeld}, \postcode{33619}, \state{NRW}, \country{Germany}}}
\affil[3]{\orgdiv{Institute for Artificial Intelligence in Medicine (IKIM)}, \orgname{University Hospital Essen, University of Duisburg-Essen}, \orgaddress{\street{Girardetstr. 2}, \city{Essen}, \postcode{45131}, \state{NRW}, \country{Germany}}}
\affil[4]{\orgdiv{Center for Virtual and Extended Reality in Medicine (ZvRM)}, \orgname{University Hospital Essen, University of Duisburg-Essen}, \orgaddress{\street{Hufelandstr. 55}, \city{Essen}, \postcode{45147}, \state{NRW}, \country{Germany}}}
\affil[5]{\orgdiv{Institute for Experimental Immunology and Imaging}, \orgname{University Hospital Essen, University of Duisburg-Essen}, \orgaddress{\street{Hufelandstr. 55}, \city{Essen}, \postcode{45147}, \state{NRW}, \country{Germany}}}

\abstract{3D data from high-resolution volumetric imaging is a central resource for diagnosis and treatment in modern medicine.
While the fast development of AI enhances imaging and analysis, commonly used visualization methods lag far behind.
Recent research used extended reality (XR) for perceiving 3D images with visual depth perception and touch but used restrictive haptic devices.
While unrestricted touch benefits volumetric data examination, implementing natural haptic interaction with XR is challenging.
The research question is whether a multisensory XR application with intuitive haptic interaction adds value and should be pursued.
In a study, 24 experts for biomedical images in research and medicine explored 3D medical shapes with 3 applications: a multisensory virtual reality (VR) prototype using haptic gloves, a simple VR prototype using controllers, and a standard PC application.
Results of standardized questionnaires showed no significant differences between all application types regarding usability and no significant difference between both VR applications regarding presence.
Participants agreed to statements that VR visualizations provide better depth information, using the hands instead of controllers simplifies data exploration, the multisensory VR prototype allows intuitive data exploration, and it is beneficial over traditional data examination methods.
While most participants mentioned manual interaction as best aspect, they also found it the most improvable.
We conclude that a multisensory XR application with improved manual interaction adds value for volumetric biomedical data examination.
We will proceed with our open-source research project ISH3DE(\textbf{I}ntuitive \textbf{S}tereoptic \textbf{H}aptic \textbf{3}D \textbf{D}ata \textbf{E}xploration) to serve medical education, therapeutic decisions, surgery preparations, or research data analysis.}


\maketitle

\section{Introduction}

Volumetric imaging techniques, e.g., Computed Tomography (CT) and Magnetic Resonance Imaging (MRI), have become indispensable resources for diagnosis and treatment in modern medicine. These advanced imaging approaches can produce high-resolution data in space and time, enabling medical professionals to obtain accurate and detailed information for clinical decision-making. Besides imaging techniques, the rapid development of artificial intelligence (AI) is another key player in the revolution of modern healthcare practices. For example, AI algorithms have enabled MRI scanning up to 10 times faster~\cite{zbontar2018fastmri}, paving the way to making portable MRI much wider accessibility. Also, AI algorithms can efficiently analyze large amounts of medical imaging data with great accuracy, as well as help medical professionals monitor patients' conditions and track treatment progress. 
However, the visual connection and interactions between volumetric imaging, AI systems, and medical professionals is a field with much-lagged advancement. 
For example, the main avenue for exploring volumetric images is still archaic, rotating pseudo-3D renderings or using artificial flythroughs with DICOM viewers on flat 2D screens.
These approaches do not support stereopsis, i.e. spatial depth perception through binocular vision~\cite{howard1995binocular}.
Nor do they support the human ability to integrate multisensory information optimally and, therefore, better perceive shapes and sizes when haptic and visual information is available~\cite{helbig2007optimal,ernst2002humans}. 
Hence, spatial features and their physical properties present in the volumetric data cannot be efficiently perceived, which can lead to overseeing important details.

As an emerging technology, extended reality (XR) can render 3D images while supporting visual depth perception.
In fact, in research, virtual reality (VR) has already been widely used for 3D volumetric medical image visualization, as reviewed in~\cite{lobachev2021inspection}, and also similarly for mixed reality~\cite{jain2023use}.
The advancement was in various directions, including new algorithms (e.g., for 3D rendering of medical data~\cite{zornack2021evaluating}), new applications (e.g., medical data on VR for educations~\cite{schloss2021uw,staubli2022magnetic}), for surgical simulations~\cite{syamlan2022virtual}, for AI-assisted surgical planning~\cite{medical_AI_VR}, for surgical planning with multi-users~\cite{chheang2021collaborative}, or new scientific discoveries (e.g., exploring human splenic microcirculation data in VR to understand how splenic red pulp capillaries join sinuses~\cite{steiniger2022human}).

Recent works~\cite{reinschluessel2021study,muender2022evaluating} have found that allowing users to touch and feel the tissues or organs displayed in VR can significantly improve the comprehensive understanding of the data.
However, to touch the virtual objects, real 3D-printed objects were used.
This is unsuitable for everyday data examination because printing volumetric data, e.g. a whole torso MRI scan, would take an extremely long time compared to data loading in XR.
Nevertheless, these works highlighted the importance of touching and manipulating beyond merely seeing. 
Since the default controllers of commercial XR sets were not explicitly designed for touching and manipulating volumetric data, people investigated how XR can be connected with additional devices for better data exploration. For example, specially designed pen-like haptic tools~\cite{faludi2019direct,zoller2020force} have been used with VR to create “haptic rendering” with better realism. Also, the combo of VR and pen-like haptic tools has been demonstrated for 3D medical image annotations to collect ground truth data for training AI models~\cite{rantamaa2023comparison}.
In specific applications, such as virtual surgical training, a pen-like device is similar to medical instruments and, therefore, could be appropriate~\cite{allgaier2022immersive}. However, in general, in volumetric data exploration, a pen-like device has limited flexibility and range of movement.
Moreover, poking surfaces with a pen is not how humans naturally explore 3D objects. 

Recently, haptic gloves have emerged as a new way to connect humans to the virtual metaverse with haptic feedback~\cite{Wang2018TWH}. 
They are lightweight wearable devices designed to provide force or tactile feedback to the fingertips and, in some cases, to the palm. 
They have sensors to track finger and palm movements with multiple degrees of freedom (DoF)~\cite{Wang2019HDF}. 
Specifically, haptic gloves enable users to interact with virtual objects in XR environments by transferring recorded finger movements to a virtual hand model. 
Whenever the virtual hand touches a virtual object, the XR application calculates haptic feedback and transmits it to the haptic gloves.
The haptic gloves, in turn, provide the feedback, thereby simulating the sensation of touching the object.

Integrating haptic gloves into an XR application for volumetric biomedical data exploration and analysis seems promising for two reasons.
First, it supports the human ability to perceive shapes and sizes through multisensory information better.
Second, this enables implementing an XR application that functions like reality and can, therefore, be operated in a completely natural and intuitive without a long learning period. 
However, the theoretically possible benefits are countered by several challenges.
Even though haptic gloves perform sufficiently for gaming and other purposes, research has shown that natural and intuitive manipulation of virtual objects is still impossible, especially if exact measures and dimensions are relevant~\cite{diss}.
Some potential reasons are identified, and fortunately, the biggest bottleneck seems to be in developing better algorithms~\cite{diss,krieger2023open}.
Nevertheless, without improving the haptic interaction with virtual objects, it is impossible to thoroughly investigate whether a multisensory XR application for volumetric biomedical image exploration and analysis is actually beneficial.
On the contrary, without the application adding value, there is no need to invest effort in improving haptic interaction.

Thus, this work aims to investigate whether it is advisable to invest in multisensory XR applications for volumetric data exploration or analysis.
To assess the potential benefits, we approached experts from our target group.
This comprises people who work professionally with volumetric images, from image creation to image analysis in research and medicine.
To enable participants to rate the usefulness of the multisensory XR application, they need to understand how different the exploration of volumetric biomedical data can be with the different application types.
This means that, at this point, the focus was not on the functional execution of a specific task for which a fully functional application would also have been needed.
The focus was instead on demonstrating the concepts, for which prototypes were also sufficient. 
Thus, we let the participants explore one volumetric data set with three applications: a standard PC application, a simple VR prototype, and a multisensory VR prototype. 
They rated their experience with standardized questionnaires and gave their opinion on the usefulness of a multisensory XR application. 

\section{Method}
\label{sec:method}

\subsection{Participants}

We enrolled 24 people (17 male, 7 female) who work professionally with volumetric (bio-) medical images, ranging from image creation to analysis.
They were recruited via mailing lists, or employees in the research institutes where it took place were approached directly and asked whether they would like to test three different software for a study and share their opinion.
Their age ranged from 23 to 65 years, with a median of 29 years.
Twenty-two participants were from the involved affiliations, but they were naive to the research project.
Nineteen participants worked in a research context only, two worked in a medical context only, and three worked within a combined context.
The study was approved by the Ethics Committee of Bielefeld University with No 2023-258.

\subsection{Procedure and Experimental Design}
Participants first signed an informed consent.
After that, they explored a volumetric data set in three applications.
That means, they had no specific task to execute, as this would require implementing software features that go beyond a simple prototype stage.
They completed a questionnaire after the usage of each application.
In the end, participants filled out a final questionnaire. 
The whole session took approximately 30 minutes.

The experimental design was a within-subjects design.
That means all participants performed all three experimental conditions corresponding to the three applications.
The first application was a \textit{standard PC application} commonly used for volumetric image visualization.
For this purpose, we chose the software ImageJ, which ran on a PC and was operated by a mouse.
It displayed the image in 2D, and to see the third dimension, the participant had to scroll through it.
The second application was the \textit{simple VR prototype} that corresponds to other VR state-of-the-art applications used in research.
This application displayed the image in 3D, and the participants could manipulate it with default VR controllers.
To \textit{grasp} virtual data, participants had to move the VR controllers close to the object and press a button. 
Once they released the button, the object was released as well.
The third application was the \textit{multisensory VR prototype}.
It differed from the simple VR prototype by allowing the participants to grasp and manipulate the virtual data intuitively using their hands.

To reduce sequence effects, the order of the three experimental conditions was varied. 
Since there are six possible orders for three experimental conditions, and we had 24 participants, each possible order occurred exactly four times.

\subsection{VR Hardware}

As VR glasses, the HTC Vive Pro 2 (HTC Corporation, Taiwan) was used to display the virtual world to the user visually.
For position tracking, we used the native system of the HTC Corporation, consisting of two base stations.
They send infrared signals detected by the HTC Vive Pro 2 and other accessories to determine position and orientation.
The HTC Vive Pro 2 was connected via cable to a PC and communicated with the server applications \textit{Vive Console} and \textit{SteamVR}.
These, in turn, communicated with the \textit{SteamVR} plugin integrated into our VR prototypes.

For the simple VR prototype, the default controllers of the HTC Vive Pro 2 were used.  
They use the same native position tracking system and communicate wirelessly with the \textit{SteamVR} applications on the PC.
Participants hold them in their hands.
When they pressed the button below their index finger, they could grasp virtual objects in the virtual world.

\begin{figure}[ht]
	\centering
    \includegraphics[width=\linewidth]{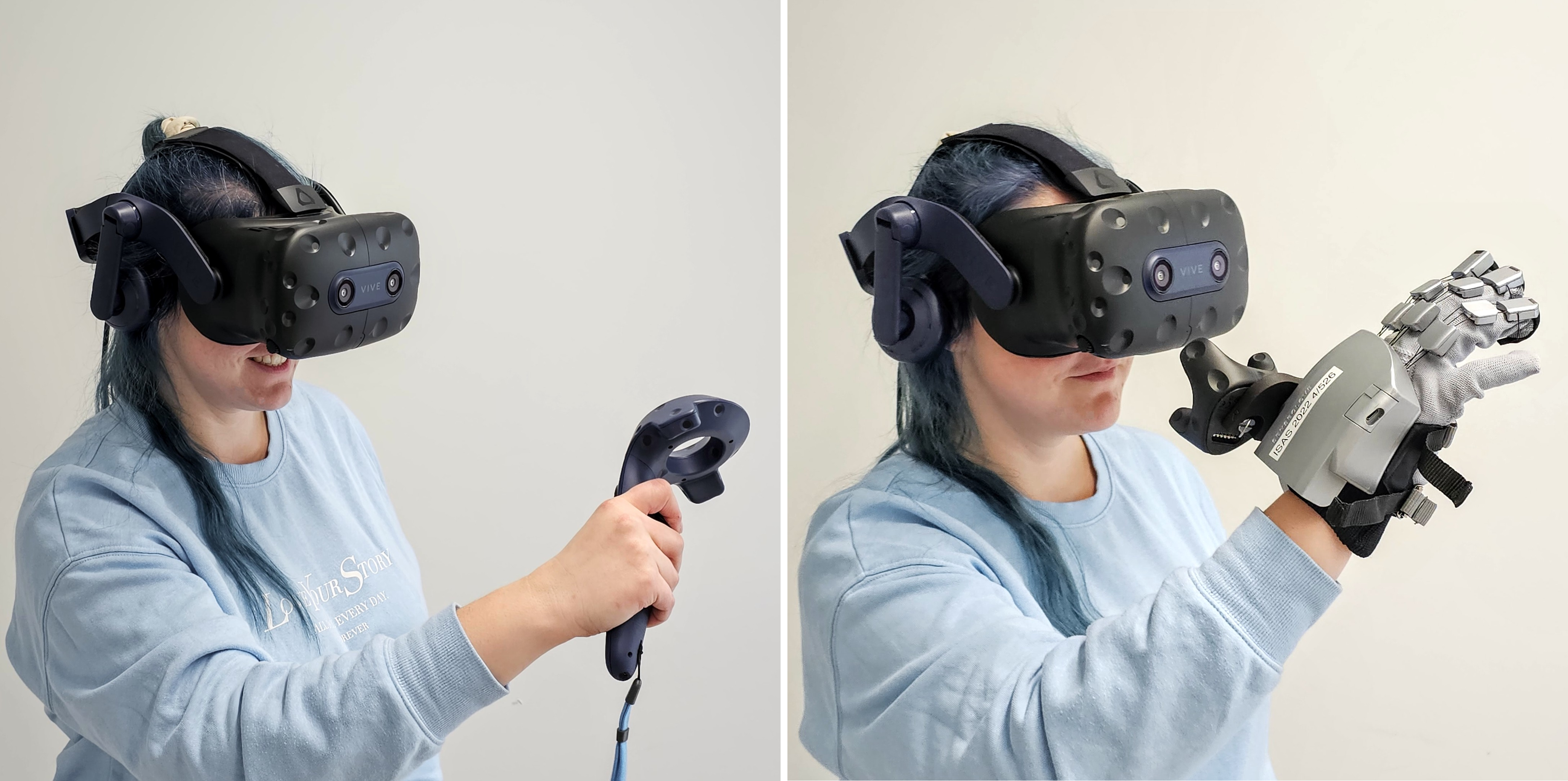}
	\caption{VR hardware. Left shows the setup used for the simple VR prototype, including HTC Vive Pro 2 and the controller. Right shows the setup used for the multisensory VR prototype, i.e. the HTC Vive Pro 2 and haptic glove SenseGlove Nova.}
	\label{fig:vr-interface}
\end{figure}

For the multisensory VR prototype, the haptic gloves SenseGlove Nova~\cite{SenseGloveNovaInfo} (SenseGlove, Netherlands) were integrated.
It comes as a pair, so that people can wear one glove on the left and one on the right hand.
It measures the finger movements with respect to the hand and sends them via Bluetooth to the server \textit{SenseCom} running on a PC.
The server, in turn, communicated with the SenseGlove plugin\footnote{The official SenseGlove software is available here \url{https://github.com/Adjuvo}.} integrated into the VR prototype.
To track the hand position and rotation in space, HTC Vive trackers were attached to them.
They also work with the HTC Corporation's native tracking system and the \textit{SteamVR} server.
Based on both sources, i.e. the hand position and orientation tracked with the HTC Vive trackers (HTC Corporation, Taiwan) and the finger movements recorded with the haptic glove, the users' detailed hand and finger movements were known.
Thus, a virtual avatar hand was displayed in VR and mimicked the users' movements in nearly real-time.
Once the user touched a virtual object, the haptic glove provided force feedback to the user's fingertips.
Figure~\ref{fig:vr-interface} shows how a user interacted with the multisensory VR prototype.

\subsection{Volumetric Medical Data}

To demonstrate the conceptual differences of the three application types to the participants, we needed a volumetric biomedical data set that could be loaded into all three applications.
We decided to use a dataset that is publicly available and can be used for the standard PC application ImageJ and VR prototypes.
Thus, we used a CT scan of a human torso.
The dataset of \textit{patient s0011} can be downloaded as a NIfTI file\footnote{\url{https://zenodo.org/records/6802614}}~\cite{Wasserthal_2023} for usage in ImageJ.
The same dataset of \textit{patient s0011} is also available, segmented and transformed into shapes, as part of MedShapeNet\footnote{\url{https://medshapenet.ikim.nrw/}}~\cite{medshapenet2023}.
MedShapeNet is a large public collection of 3D medical shapes, e.g., bones, organs, and vessels.
They are based on original data from various databases, mostly captured with medical imaging devices such as computed tomography (CT) and magnetic resonance imaging (MRI) scanners.
Those data were already processed, i.e. segmented with different AI methods, converted to shapes, and labeled with the name of the corresponding anatomy.
These shapes consist of triangular meshes and point clouds and can be downloaded in \textit{.stl} format.
Thus, they can be easily loaded into VR applications where they have the exact same size as the scanned human torso.

\begin{figure*}[ht]
	\centering
    \includegraphics{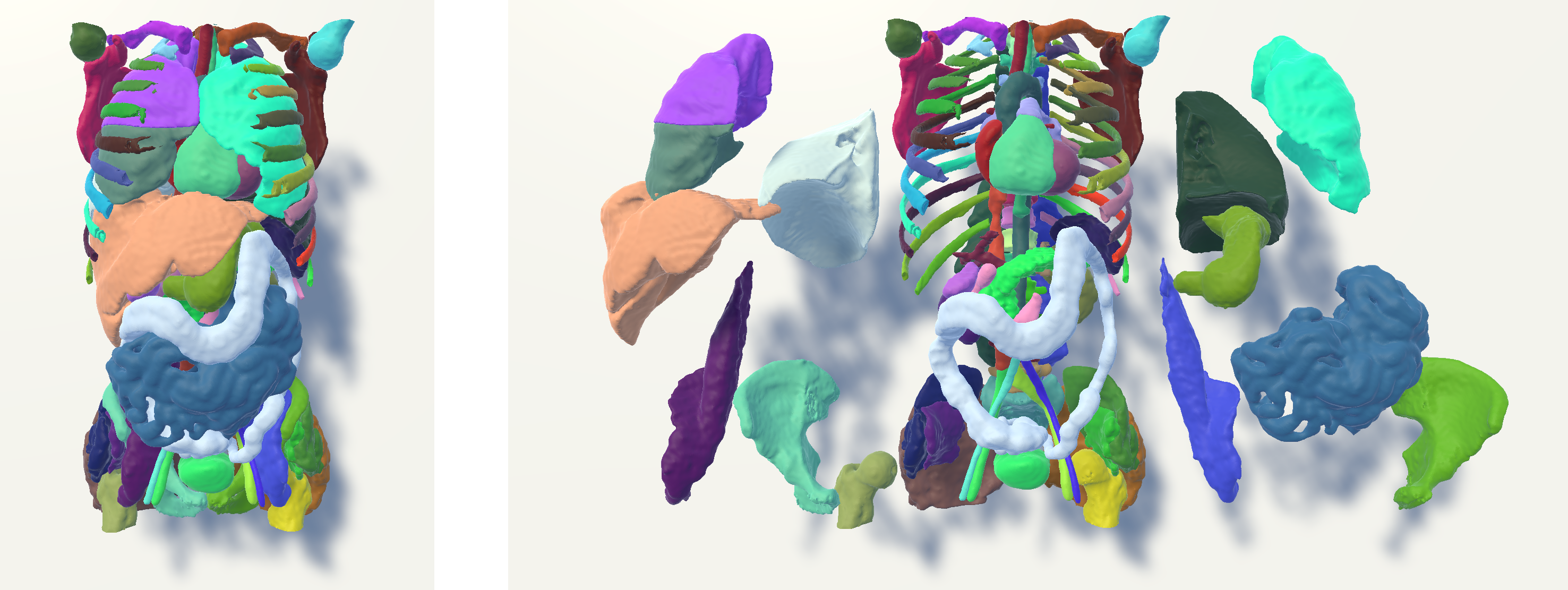}
	\caption{Torso data of patient s0011 from the MedShapeNet data set (\url{https://medshapenet.ikim.nrw/}). \textit{Left} shows how the data was originally loaded. \textit{Right} shows how the upper body could be disassembled to examine how individual organs are located in the body.}
	\label{fig:body}
\end{figure*}

Summarizing, a CT human torso scan as an original NIfTI file was loaded into ImageJ, while the same human torso segmented into 104 parts, e.g. representing bones, organs, and vessels, was used for the VR prototypes. The latter one is shown in Fig~\ref{fig:body}.

\subsection{The VR Prototypes}
Both prototypes were implemented in C\# with Unity and consisted of a VR scene and several modules.
For each hardware component, a module handled the communication with the device and its integration into the VR scene.
One module contained the virtual environment, which was a virtual room in which the user could walk around and manipulate the data.
Another module handled the volumetric data.
The volumetric data parts were set up to be grasped and manipulated through the haptic gloves or the VR controllers using the libraries provided with the hardware.

\begin{figure}[ht]
	\centering
 \includegraphics{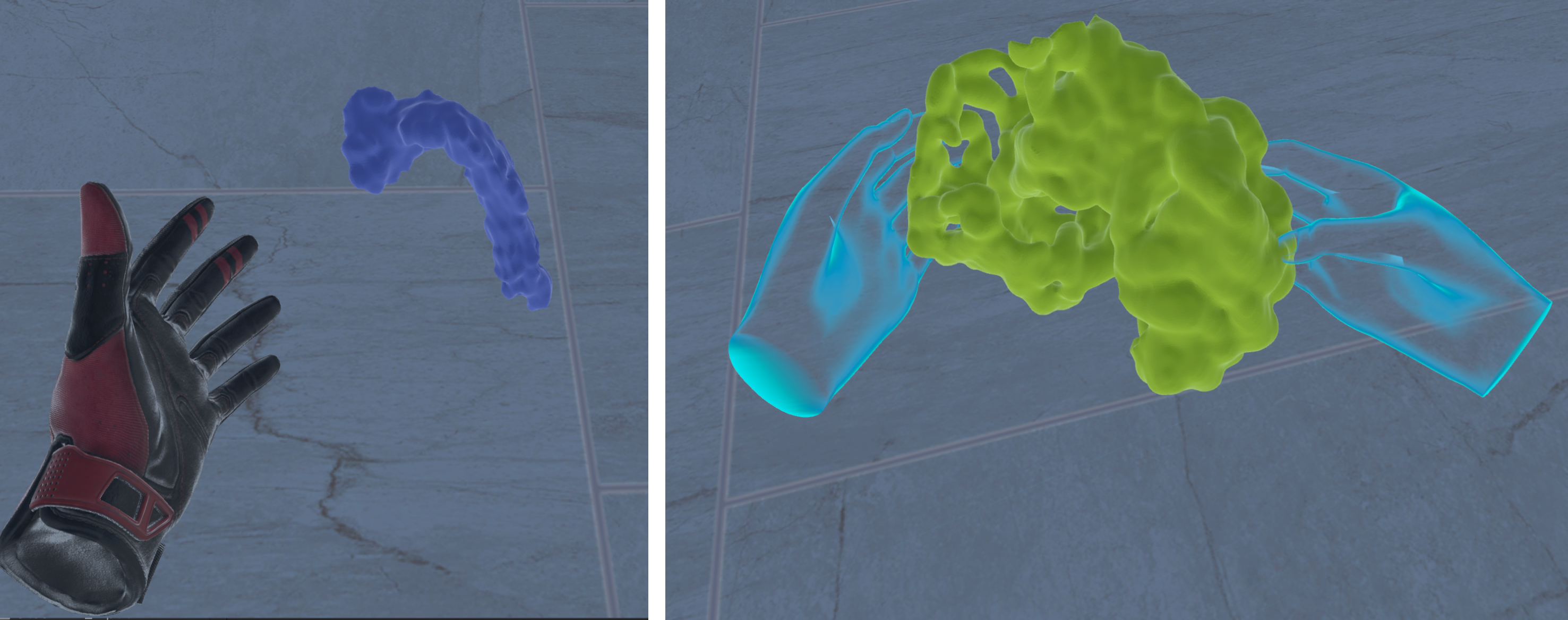}
	\caption{A participant grasps a 3D medical shape from MedShapeNet. Left shows using the simple VR prototype with the controllers and the native SteamVR library which makes the hands disappearing during grasping. Right shows using the multisensory VR prototype with the native SenseGlove library.}
	\label{fig:hand-grasp-stomach}
\end{figure}

After starting the VR prototypes, participants were in a room where they could move freely. 
In the middle of the room, the whole torso scan was displayed, see Figure~\ref{fig:body} left.
In the simple VR prototype, participants saw their virtual hands as red and black gloves corresponding to the \textit{SteamVR} plugin.
By pressing buttons on the VR controller, they could manipulate the 3D medical shapes.
In the multisensory VR prototype, participants' virtual hands were displayed as blue transparent meshes based on the original plugin of SenseGlove, shown in Fig.~\ref{fig:hand-grasp-stomach}.
The virtual hand moved according to the participants' hand movements.
Participants could manipulate the 3D medical shapes using their hands as they would manipulate objects in reality.
In both applications, the possible manipulations of the 3D medical shapes were to grasp, move, and rotate them for visual inspection from different angles.
Moreover, participants could examine how the organs were located by disassembling the torso organ-wise, see Figure~\ref{fig:body} right.

\subsection{Measures}
To assess participants' experience with the three application types and to allow for objective comparison, participants filled out standardized, established questionnaires immediately after trying each application.
To check if participants found the application usable, we chose the standardized \textit{System Usability Scale (SUS)}~\cite{Brooke1995SUS}, which provides insights into usability.
It had ten questions and was answered on a 5-point Likert scale.
Based on a given formula~\cite{Brooke1995SUS}, the ten answers were added together, resulting in a SUS score of 0 (poor usability) to 100 (perfect usability) per participant.
Further, we intended to check whether participants could do the volumetric data exploration as they wished.
Since, especially in VR applications, the ability to \textit{do} impacts the feeling of presence~\cite{Sanchez2005presence}, we chose as the second questionnaire the \textit{IGroup Presence Questionnaire (IPQ)}~\cite{Schubert2001experience}.
The IPQ consists of four subcategories: the general presence \textit{G}, the spatial presence \textit{SP}, the involvement \textit{INV}, and the experienced realism \textit{REAL}.
Thereby, involvement describes how strongly one is focused on interacting with the application.
While \textit{G} consists of one question, the other subcategories consist of multiple questions.
All questions were answered on a 7-point Likert scale.
They were added together based on a formula~\cite{Schubert2001experience}.
Thus the scores for the subcategories range from 0 (not evident) to 6 (very prominent).

After trying all application types and filling out the corresponding questionnaires, participants completed a final questionnaire.
The goal was to investigate whether a multisensory XR application adds value for volumetric biomedical data exploration or analysis in a medical or research context.
It contained four subjective questions regarding the intentions of creating a multisensory XR application, which had to be answered on a 5-point Likert scale and are listed in Table~\ref{tab:questions}.
Moreover, they were asked two open questions on what they liked the most and what we must improve the most about the multisensory XR application.
Finally, it also contained a question about the gender and the profession.

\section{Results}
\label{sec:results}
The results of the SUS score are in Table~\ref{tab:sus-results}.
The ratings of the usability were, on average, similar for all three application types.
As the SUS score ranged between 0 and 100, we chose an ANOVA with repeated measures for comparison.
It did not reveal any significant differences with $F(38,2)=0.39, p=0.677$.

\begin{table*}[h]
\caption{Results of the standardized SUS questionnaire. The multisensory VR prototype is abbreviated with \textit{multi}, the simple VR prototype with \textit{simple}, and the standard PC application with \textit{standard}.}
\label{tab:sus-results}
\centering
\begin{tabularx}{84mm}{lXXX}
\toprule
&multi&simple&standard\\
\midrule
$M$&73.00&76.25&75.63\\
$SD$&12.61&14.20&18.17\\
\botrule
&&&\\
\end{tabularx}
\\
\includegraphics{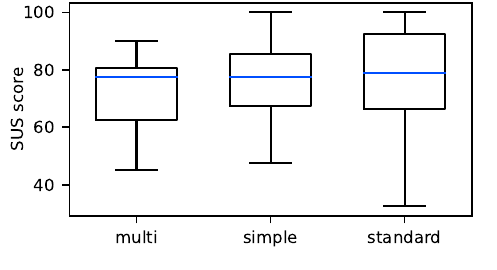}
\end{table*}

Participants filled the IPQ questionnaire for all applications, to keep the experimental design simple and the participants unbiased.
However, the questionnaire was created for VR applications, and some questions do not fit to describe the experience with a PC application.
Therefore, it seems reasonable to exclude the IPQ results for the standard PC application.
Table~\ref{tab:standardized-results} shows the mean and standard deviation over all participants for all IPQ subcategories for the multisensory VR prototype and the simple VR prototype.
The mean values for all subcategories were similar for both applications.
Since the first subcategory of IPQ consisted just of one question and those were answered on a 7-point Likert scale, we compared the results with a non-parametric test, i.e. the Wilcoxon-signed-rank test.
There are no significant differences between both VR prototypes.
This means we found no significant differences of how present in general, how present in space, how involved, and how realistic participants perceived both VR prototypes.
The results are displayed in Table~\ref{tab:standardized-results} in the box plot figures.

\begin{table*}[h]
\caption{Results of the standardized IPQ questionnaire. The multisensory VR prototype is abbreviated with \textit{multi}, and the simple VR prototype with \textit{simple}. Both applications were compared with a Wilcoxon-signed-rank test.}
\label{tab:standardized-results}
\scriptsize
\begin{tblr}{width=130mm, colspec={X[1,l]X[1,c]X[1,c]X[1,c]X[1,c]X[1,c]X[1,c]X[1,c]X[1,c]}}
\hline
& \SetCell[c=2]{c}{{{G}}}&& \SetCell[c=2]{c}{{{SP}}} && \SetCell[c=2]{c}{{{INV}}} && \SetCell[c=2]{c}{{{REAL}}}\\
&multi&simple&multi&simple&multi&simple&multi&simple\\
\hline
$M$&4.83&4.54&4.70&4.50&3.29&3.27&2.30&2.21\\
$SD$&1.20&1.35&1.27&1.18&1.26&1.23&1.11&1.05\\
\hline
$W$& \SetCell[c=2]{c}{{{21.0}}}&& \SetCell[c=2]{c}{{{65.5}}} && \SetCell[c=2]{c}{{{93.0}}} && \SetCell[c=2]{c}{{{87.5}}}\\
$p$& \SetCell[c=2]{c}{{{$0.271$}}}&& \SetCell[c=2]{c}{{{$0.383$}}} && \SetCell[c=2]{c}{{{$0.936$}}} && \SetCell[c=2]{c}{{{$0.511$}}}\\
\hline
\end{tblr}
\begin{tblr}{width=130mm, colspec={X[1,l]X[2,c]X[2,c]X[2,c]X[2,c]}}
\rotatebox[origin=c]{90}{Scores}
&
 \begin{minipage}{\linewidth}
      \includegraphics[width=\linewidth]{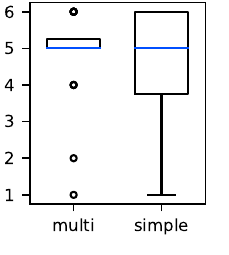}
\end{minipage}
&
 \begin{minipage}{\linewidth}
      \includegraphics[width=\linewidth]{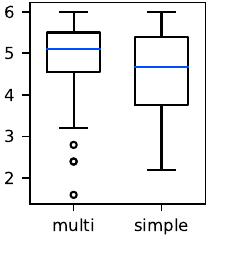}
\end{minipage}
&
 \begin{minipage}{\linewidth}
      \includegraphics[width=\linewidth]{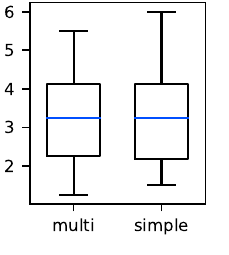}
\end{minipage}
&
 \begin{minipage}{\linewidth}
      \includegraphics[width=\linewidth]{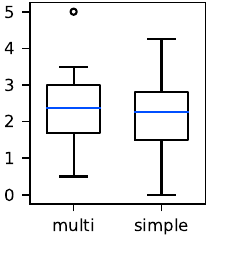}
\end{minipage}
\end{tblr}
\end{table*}

The results of the final questionnaire are summarized in Table~\ref{tab:questions}.
Since the possible answers were from 1 (strongly disagree) to 5 (strongly agree), a value of 3 would indicate a neutral answer.
The mean and median values are, on average, higher than 3.
We wanted to examine whether the distribution of answers differs significantly from disagreement or a neutral answer.
So, we performed a one-sided, one-sampled t-tests against 3 with $\alpha=0.05$.
As the Table~\ref{tab:questions} shows, all t-tests delivered significant results.
This means the distribution of answers differs significantly from disagreement.
Thus, overall, participants seemed to agree with the statements.

\begin{table*}[ht]
\begin{center}
\caption{Questions of the self-administered questionnaire answered on a Likert scale of 1 (strongly disagree) to 5 (strongly agree), and their results. In this context, \textit{ISH3DE} was introduced as the name for the multisensory VR prototype.}
\label{tab:questions}
\scriptsize
\begin{tblr}{width=130mm, colspec={X[1,l]X[3,h]X[3,h]X[3,h]X[3,h]}}
&\centering\textbf{Q1}&\centering\textbf{Q2}&\centering\textbf{Q3}&\centering\textbf{Q4}\\
\hline
&
I think a data visualization in VR helped me get a better understanding of the organs' sizes and dimensions than a  visualization on a PC screen.&
I think using the hands for exploring the data instead of a VR controller simplifies the data exploration.&
Exploring the data with ISH3DE was intuitive.&
Overall, I believe using ISH3DE for volumetric medical data examination is beneficial over the traditional method.\\
\hline
$Mdn$&\centering5&\centering3.5&\centering4&\centering4\\
$M$&\centering4.7&\centering3.5&\centering4&\centering3.8\\
$SD$&\centering0.54&\centering1.22&\centering0.87&\centering1.21\\
\hline
$t(23)$&\centering15.22&\centering1.96&\centering5.54&\centering3.29\\
$p$&\centering$<0.001$&\centering$0.031$&\centering$<0.001$&\centering$0.002$\\
\hline
\rotatebox[origin=c]{90}{Frequency}
&
 \begin{minipage}{\linewidth}
      \includegraphics[width=\linewidth]{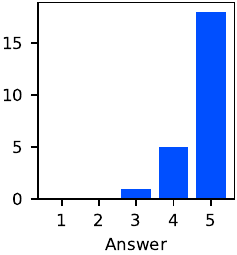}
\end{minipage}
&
 \begin{minipage}{\linewidth}
      \includegraphics[width=\linewidth]{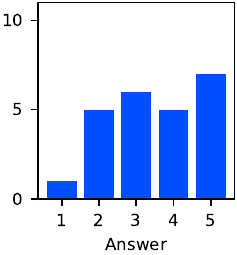}
\end{minipage}
&
 \begin{minipage}{\linewidth}
      \includegraphics[width=\linewidth]{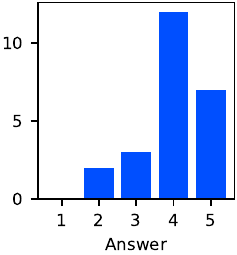}
\end{minipage}
&
 \begin{minipage}{\linewidth}
      \includegraphics[width=\linewidth]{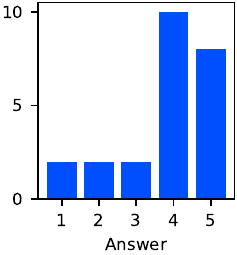}
\end{minipage}
\\
\end{tblr}
\end{center}
\end{table*}
For the open questions, participants stated that they liked most: usage and freedom of the hands (instead of VR controllers), haptic or force feedback and sense of touch, visualization of the organs, better perceiving the organs' dimensions and shapes, feeling more \textit{inside} reality while being in VR or being immersed, easy to put on, more natural interaction or intuitive usage, ability to rotate and inspect the organs, simplicity.
Based on the participants, the most improvable things were:
the hardware reacted jerkily or jumped, the haptic or force feedback (should be more smoothly, more realistic, and have better timing), the accuracy of grasping (i.e. a gap between fingers and grasped objects, issues with edges, sometimes wrong organs are grasped), system robustness, too high force or resistance, grasping partially hidden objects (e.g. by using a ray cast feature), commonly used navigation through recorded slices is missing, details or structure of organs is missing, a mismatch between real and virtual hand movement, adding texture corresponding to the actual organ, more lightweight haptic gloves.

\section{Discussions}
In this work, we investigated whether a multisensory XR application can add value for volumetric biomedical data examination in medicine and research.
For this purpose, we conducted a study with 24 participants who work professionally with biomedical images in research and medical contexts.
The experts tested a multisensory VR prototype in which data could be manipulated intuitively using the hands.
In comparison, they used a simple VR prototype that uses VR controllers for data manipulation and is comparable with state-of-the-art research, as well as a standard PC application for volumetric data examination.

Results of a standardized usability questionnaire showed no significant differences between the three application types. 
The presence questionnaire revealed no significant differences between the two VR applications.
So, based on the standardized usability or presence questionnaires, the multisensory VR application neither adds value nor is disadvantageous for volumetric biomedical data exploration and analysis.

However, the final questionnaire asked the participants for their opinions on several statements.
The outcome shows that participants think that a multisensory XR application helps to get a better understanding of the organs' dimensions.
They agreed that using hands instead of VR controllers simplifies data exploration.
They felt that the multisensory VR prototype allowed intuitive data exploration.
Finally, they believed that a multisensory XR application might be beneficial over traditional data exploration methods.
So, based on this final questionnaire outcome, it seems that a multisensory XR application does add value for biomedical data examination.

The study had, however, several limitations.
Despite the possibility to manually interact with the volumetric data was mentioned most often as one of the most positive aspects of the multisensory VR prototype, it was also listed most often as the most improvable aspect. 
The bottleneck of the current state of haptic gloves was already known in advance.
Currently, most haptic gloves available on the market provide just heuristics~\cite{krieger2023open} instead of accurate hand postures leading to a mismatch between actual hand movements and the corresponding simulation in VR.
Therefore, we assume that for a fully working VR application, we have to work on improving haptic algorithms.
This means, in the current study, participants did not test the multisensory VR application with the haptic perception as we intend it for the long run.
However, considering the problems of manual interaction, it is impressive that the results of the presence questionnaire for the multisensory VR prototype were as good as those for the simple VR prototype. 
The same holds for the results of the usability questionnaire.
We would argue that an improved manual interaction might even lead to better outcomes in the standardized questionnaires. 

Another study limitation is the fact, that the participants had no explicit task during the study and just explored the data.
In order to evaluate the usability of such an application it would be required to perform explicit and real-life tasks with different applications.
However, the VR applications are currently just in a prototype stage.
Therefore, it was not possible to perform real life tasks, e.g. diagnosing a tumor, with those prototypes yet.
Implementing all those features, in order to allow real-life tasks, would have been an immense task. 
And first completing this immense task and perform a first evaluation just thereafter seemed not a good strategy. 
The reason is, that before testing, it was unclear, whether people would like such a multisensory VR prototype for those tasks or prefer the standard PC application anyway.
So in the worst case, this creating a fully working application would have been a time waste.
Therefore, this current, very limited study just shows what the participants \textit{think} such an application may or may not offer.
Of course, it is required to perform a usability studies with explicit real-life tasks in a later development stage of the project in order to provide actual statements about the usefulness.

In conclusion, based on the questionnaire results being equally good despite the problematic manual interaction and especially based on the experts' opinions, our proposed multisensory XR application for biomedical data exploration and analysis might add value. 
Thus, we aim to proceed with working on this research project.

\section{Future work}
\begin{figure*}
	\centering
	\includegraphics[width=\linewidth]{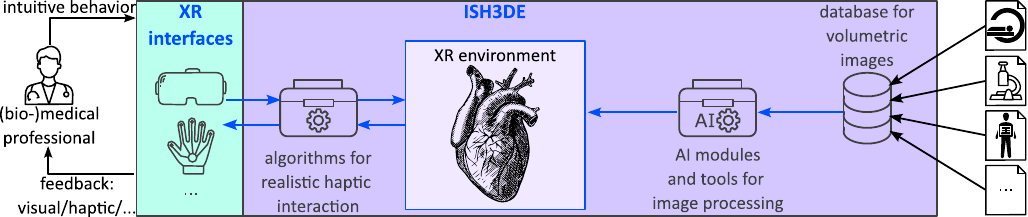}
	\caption{Schematic representation of ISH3DE.}
	\label{fig:concept}
\end{figure*}

We intend to further work on our research project that we named ISH3DE, a framework for \textbf{I}ntuitive \textbf{S}tereoptic \textbf{H}aptic \textbf{3}D \textbf{D}ata \textbf{E}xploration.
The overall concept of our proposed ISH3DE is illustrated in Figure~\ref{fig:concept}, where the information can freely flow through the (bio-) medical professionals, the AI system, and the volumetric data via interactions in an XR scene. 
The information flow between volumetric data and (bio-) medical professionals, now called \textit{users}, passes through the XR interface consisting of multiple hardware components.
It allows the users to intuitively interact with rendered data parts, e.g. organs, by grasping, feeling, and inspecting them from different perspectives.
The data flow between the users and the AI systems runs through a virtual graphical user interface in the XR scene.
Users can send a request to the AI system for various actions (e.g. to enhance the image quality or segment the organs), which are then applied to the rendered volumetric data.
In addition, users can also provide feedback to the AI system to further train it, such as collecting a preliminary diagnosis label.
The framework code is released open-source here (\url{https://github.com/MMV-Lab/ISH3DE}).

To improve the manual interaction, we will base on our previous work, in which a hand calibration tool was implemented to create an adequate hand model that fits the user's hand~\cite{krieger2023open}.
Some of our on-going projects include an algorithm calculating the hand posture based on fingertip positions so that the haptic glove can measure and provides more accurate hand posture in VR.
Additionally, the haptic feedback will be improved by rendering different elasticities of objects and enabling deformability.
For this purpose, Young's modulus of common soft tissues or organs~\cite{liu2015hydrogels} can be encoded.

A limitation of the current prototype is that the interaction between the ISH3DE and AI systems is still offline.
In other words, the currently loaded data is pre-processed by AI models and converted into proper binary and ASCII \textit{.stl} files. 
The next step of our work is to utilize ONNX~\cite{bai2019} and Unity neural network inference engine (Barracuda) to hook up multiple AI models to the XR application.
This will allow the loading of raw volumetric data and request different AI models for different tasks, e.g. segmentation or labeling.

Another essential part of future work is close cooperation with the target groups from medicine and research.
In particular, before the software content is further developed, a focus group should work out requirements.
In addition, studies should be conducted in which real use cases are tested.
The performance between ISH3DE and traditional software
can be compared in such downstream tasks. 

\section{Conclusion}
We presented a new viable path for medical professionals or researchers to explore and analyze volumetric biomedical data.
It is based on the premise that supporting the human's ability of multisensory depth perception delivers a better understanding of the data's dimensions.
A user study demonstrated that experts agree this new data exploration paradigm adds value.
Therefore, we will proceed to work on the project ISH3DE, which is also open source.
In the future, it may lead to more effective therapeutic decisions, better surgery preparations, enhanced medical education, or improved research data analysis.

\backmatter

\section*{Declarations}
\bmhead{Funding}
This work was supported by KITE (Plattform für KI-Trans\-lation Essen) from the REACT-EU initiative (EFRE-0801977, \url{https://kite.ikim.nrw/}).
J.Chen is funded by the Federal Ministry of Education and Research (Bundesministerium f\"ur Bildung und Forschung, BMBF) in Germany under the funding reference 161L0272. 
J. Chen, K. Krieger, M. Gunzer are also supported by the Ministry of Culture and Science of the State of North Rhine-Westphalia (Ministerium f\"ur Kultur und Wissenschaft des Landes Nordrhein-Westfalen, MKW NRW).

\bmhead{Competing interests}
The authors have no relevant financial or non-financial interests to disclose.

\bmhead{Ethics approval}
This study was performed in line with the principles of the German Society of Psychology (Deutschen Gesellschaft für Psychologie, DGPs) and the Professional Association of German Psychologists (Berufsverbandes deutscher Psychologinnen und Psychologen, BdP). 
The Ethics Committee of Bielefeld University granted approval, with No 2023-258, on September 12th, 2023. 

\bmhead{Consent to participate}
Informed consent was obtained from all individual participants included in the study.

\bmhead{Consent for publication}
All participants agreed that their data could be published anonymously.

\bmhead{Code availability}
The framework code that was used for the study and all further developments are released open-source here (\url{https://github.com/MMV-Lab/ISH3DE}).

\bmhead{Author contribution}
Conceptualization of the research was performed by Kathrin Krieger, Matthias Gunzer, and Jianxu Chen. 
Project administration, software creation, study conceptualization, design and execution, data analysis, and visualization were performed by Kathrin Krieger.
Resources and funding acquisition were organized by Jan Egger, Jens Kleesiek, Matthias Gunzer, and Jianxu Chen.
The first draft of the manuscript was written by Kathrin Krieger and Jianxu Chen, and all authors commented on previous versions of the manuscript. 
All authors read and approved the final manuscript.

\bigskip


\bibliography{mybibliography}

\begin{thebibliography}{31}
\expandafter\ifx\csname natexlab\endcsname\relax\def\natexlab#1{#1}\fi
\providecommand{\url}[1]{\texttt{#1}}
\providecommand{\href}[2]{#2}
\providecommand{\path}[1]{#1}
\providecommand{\DOIprefix}{doi:}
\providecommand{\ArXivprefix}{arXiv:}
\providecommand{\URLprefix}{URL: }
\providecommand{\Pubmedprefix}{pmid:}
\providecommand{\doi}[1]{\href{http://dx.doi.org/#1}{\path{#1}}}
\providecommand{\Pubmed}[1]{\href{pmid:#1}{\path{#1}}}
\providecommand{\bibinfo}[2]{#2}
\ifx\xfnm\relax \def\xfnm[#1]{\unskip,\space#1}\fi
\bibitem[{Allgaier et~al.(2022)Allgaier, Neyazi, Sandalcioglu, Preim and
  Saalfeld}]{allgaier2022immersive}
\bibinfo{author}{Allgaier, M.}, \bibinfo{author}{Neyazi, B.},
  \bibinfo{author}{Sandalcioglu, I.E.}, \bibinfo{author}{Preim, B.},
  \bibinfo{author}{Saalfeld, S.}, \bibinfo{year}{2022}.
\newblock \bibinfo{title}{Immersive vr training system for clipping
  intracranial aneurysms}.
\newblock \bibinfo{journal}{Current Directions in Biomedical Engineering}
  \bibinfo{volume}{8}, \bibinfo{pages}{9--12}.
\bibitem[{Bai et~al.(2019)Bai, Lu, Zhang et~al.}]{bai2019}
\bibinfo{author}{Bai, J.}, \bibinfo{author}{Lu, F.}, \bibinfo{author}{Zhang,
  K.}, et~al., \bibinfo{year}{2019}.
\newblock \bibinfo{title}{Onnx: Open neural network exchange}.
\newblock \bibinfo{howpublished}{\url{https://github.com/onnx/onnx}}.
\bibitem[{Brooke(1995)}]{Brooke1995SUS}
\bibinfo{author}{Brooke, J.}, \bibinfo{year}{1995}.
\newblock \bibinfo{title}{Sus: A quick and dirty usability scale}.
\newblock \bibinfo{journal}{Usability Eval. Ind.} \bibinfo{volume}{189}.
\bibitem[{Chheang et~al.(2021)Chheang, Saalfeld, Joeres, Boedecker, Huber,
  Huettl, Lang, Preim and Hansen}]{chheang2021collaborative}
\bibinfo{author}{Chheang, V.}, \bibinfo{author}{Saalfeld, P.},
  \bibinfo{author}{Joeres, F.}, \bibinfo{author}{Boedecker, C.},
  \bibinfo{author}{Huber, T.}, \bibinfo{author}{Huettl, F.},
  \bibinfo{author}{Lang, H.}, \bibinfo{author}{Preim, B.},
  \bibinfo{author}{Hansen, C.}, \bibinfo{year}{2021}.
\newblock \bibinfo{title}{A collaborative virtual reality environment for liver
  surgery planning}.
\newblock \bibinfo{journal}{Computers \& Graphics} \bibinfo{volume}{99},
  \bibinfo{pages}{234--246}.
\bibitem[{Ernst and Banks(2002)}]{ernst2002humans}
\bibinfo{author}{Ernst, M.O.}, \bibinfo{author}{Banks, M.S.},
  \bibinfo{year}{2002}.
\newblock \bibinfo{title}{Humans integrate visual and haptic information in a
  statistically optimal fashion}.
\newblock \bibinfo{journal}{Nature} \bibinfo{volume}{415},
  \bibinfo{pages}{429--433}.
\bibitem[{Faludi et~al.(2019)Faludi, Zoller, Gerig, Zam, Rauter and
  Cattin}]{faludi2019direct}
\bibinfo{author}{Faludi, B.}, \bibinfo{author}{Zoller, E.I.},
  \bibinfo{author}{Gerig, N.}, \bibinfo{author}{Zam, A.},
  \bibinfo{author}{Rauter, G.}, \bibinfo{author}{Cattin, P.C.},
  \bibinfo{year}{2019}.
\newblock \bibinfo{title}{Direct visual and haptic volume rendering of medical
  data sets for an immersive exploration in virtual reality}, in:
  \bibinfo{booktitle}{Medical Image Computing and Computer Assisted
  Intervention--MICCAI 2019: 22nd International Conference, Shenzhen, China,
  October 13--17, 2019, Proceedings, Part V 22},
  \bibinfo{organization}{Springer}. pp. \bibinfo{pages}{29--37}.
\bibitem[{Helbig and Ernst(2007)}]{helbig2007optimal}
\bibinfo{author}{Helbig, H.B.}, \bibinfo{author}{Ernst, M.O.},
  \bibinfo{year}{2007}.
\newblock \bibinfo{title}{Optimal integration of shape information from vision
  and touch}.
\newblock \bibinfo{journal}{Experimental brain research} \bibinfo{volume}{179},
  \bibinfo{pages}{595--606}.
\bibitem[{Howard and Rogers(1995)}]{howard1995binocular}
\bibinfo{author}{Howard, I.P.}, \bibinfo{author}{Rogers, B.J.},
  \bibinfo{year}{1995}.
\newblock \bibinfo{title}{Binocular vision and stereopsis}.
\newblock \bibinfo{publisher}{Oxford University Press, USA}.
\bibitem[{Jain et~al.(2023)Jain, Gao, Yeo and Ngiam}]{jain2023use}
\bibinfo{author}{Jain, S.}, \bibinfo{author}{Gao, Y.}, \bibinfo{author}{Yeo,
  T.T.}, \bibinfo{author}{Ngiam, K.Y.}, \bibinfo{year}{2023}.
\newblock \bibinfo{title}{Use of mixed reality in neuro-oncology: A single
  centre experience}.
\newblock \bibinfo{journal}{Life} \bibinfo{volume}{13}, \bibinfo{pages}{398}.
\bibitem[{Krieger(submitted)}]{diss}
\bibinfo{author}{Krieger, K.}, \bibinfo{year}{submitted}.
\newblock \bibinfo{title}{A VR Serious Game Framework for Haptic Performance
  Evaluation}.
\newblock Ph.D. thesis. Bielefeld University.
\bibitem[{Krieger et~al.(2023)Krieger, Leins, Markmann and
  Haschke}]{krieger2023open}
\bibinfo{author}{Krieger, K.}, \bibinfo{author}{Leins, D.P.},
  \bibinfo{author}{Markmann, T.}, \bibinfo{author}{Haschke, R.},
  \bibinfo{year}{2023}.
\newblock \bibinfo{title}{Open-source hand model configuration tool (hmct)}.
\newblock \bibinfo{journal}{Work-in-Progress Paper at 2023 IEEE Worldhaptics
  Conference} .
\bibitem[{Krieger et~al.(submitted)Krieger, Leins, Markmann, Haschke, Chen,
  Gunzer and Ritter}]{Krieger2023AKM}
\bibinfo{author}{Krieger, K.}, \bibinfo{author}{Leins, D.P.},
  \bibinfo{author}{Markmann, T.}, \bibinfo{author}{Haschke, R.},
  \bibinfo{author}{Chen, J.}, \bibinfo{author}{Gunzer, M.},
  \bibinfo{author}{Ritter, H.}, \bibinfo{year}{submitted}.
\newblock \bibinfo{title}{Adaptive kinematic modeling for improved hand posture
  estimates using a haptic glove}.
\newblock \bibinfo{journal}{.} .
\bibitem[{Li et~al.(2023)Li, Pepe, Gsaxner et~al.}]{medshapenet2023}
\bibinfo{author}{Li, J.}, \bibinfo{author}{Pepe, A.}, \bibinfo{author}{Gsaxner,
  C.}, et~al., \bibinfo{year}{2023}.
\newblock \bibinfo{title}{Medshapenet - a large-scale dataset of 3d medical
  shapes for computer vision}.
\newblock \bibinfo{journal}{arXiv preprint arXiv:2308.16139} .
\bibitem[{Liu et~al.(2015)Liu, Zheng, Poh, Machens and
  Schilling}]{liu2015hydrogels}
\bibinfo{author}{Liu, J.}, \bibinfo{author}{Zheng, H.}, \bibinfo{author}{Poh,
  P.S.}, \bibinfo{author}{Machens, H.G.}, \bibinfo{author}{Schilling, A.F.},
  \bibinfo{year}{2015}.
\newblock \bibinfo{title}{Hydrogels for engineering of perfusable vascular
  networks}.
\newblock \bibinfo{journal}{International journal of molecular sciences}
  \bibinfo{volume}{16}, \bibinfo{pages}{15997--16016}.
\bibitem[{Lobachev et~al.(2021)Lobachev, Berthold, Pfeffer, Guthe and
  Steiniger}]{lobachev2021inspection}
\bibinfo{author}{Lobachev, O.}, \bibinfo{author}{Berthold, M.},
  \bibinfo{author}{Pfeffer, H.}, \bibinfo{author}{Guthe, M.},
  \bibinfo{author}{Steiniger, B.S.}, \bibinfo{year}{2021}.
\newblock \bibinfo{title}{Inspection of histological 3d reconstructions in
  virtual reality}.
\newblock \bibinfo{journal}{Frontiers in Virtual Reality} \bibinfo{volume}{2},
  \bibinfo{pages}{628449}.
\bibitem[{Muender et~al.(2022)Muender, Reinschluessel, Salzmann, L{\"u}ck,
  Schenk, Weyhe, D{\"o}ring and Malaka}]{muender2022evaluating}
\bibinfo{author}{Muender, T.}, \bibinfo{author}{Reinschluessel, A.V.},
  \bibinfo{author}{Salzmann, D.}, \bibinfo{author}{L{\"u}ck, T.},
  \bibinfo{author}{Schenk, A.}, \bibinfo{author}{Weyhe, D.},
  \bibinfo{author}{D{\"o}ring, T.}, \bibinfo{author}{Malaka, R.},
  \bibinfo{year}{2022}.
\newblock \bibinfo{title}{Evaluating soft organ-shaped tangibles for medical
  virtual reality}, in: \bibinfo{booktitle}{CHI Conference on Human Factors in
  Computing Systems Extended Abstracts}, pp. \bibinfo{pages}{1--8}.
\bibitem[{Rantamaa et~al.(2023)Rantamaa, Kangas, Kumar, Mehtonen, J{\"a}rnstedt
  and Raisamo}]{rantamaa2023comparison}
\bibinfo{author}{Rantamaa, H.R.}, \bibinfo{author}{Kangas, J.},
  \bibinfo{author}{Kumar, S.K.}, \bibinfo{author}{Mehtonen, H.},
  \bibinfo{author}{J{\"a}rnstedt, J.}, \bibinfo{author}{Raisamo, R.},
  \bibinfo{year}{2023}.
\newblock \bibinfo{title}{Comparison of a vr stylus with a controller, hand
  tracking, and a mouse for object manipulation and medical marking tasks in
  virtual reality}.
\newblock \bibinfo{journal}{Applied Sciences} \bibinfo{volume}{13},
  \bibinfo{pages}{2251}.
\bibitem[{Reinschluessel et~al.(2021)Reinschluessel, Muender, D{\"o}ring,
  Uslar, L{\"u}ck, Weyhe, Schenk and Malaka}]{reinschluessel2021study}
\bibinfo{author}{Reinschluessel, A.V.}, \bibinfo{author}{Muender, T.},
  \bibinfo{author}{D{\"o}ring, T.}, \bibinfo{author}{Uslar, V.N.},
  \bibinfo{author}{L{\"u}ck, T.}, \bibinfo{author}{Weyhe, D.},
  \bibinfo{author}{Schenk, A.}, \bibinfo{author}{Malaka, R.},
  \bibinfo{year}{2021}.
\newblock \bibinfo{title}{A study on the size of tangible organ-shaped
  controllers for exploring medical data in vr}, in:
  \bibinfo{booktitle}{Extended Abstracts of the 2021 CHI Conference on Human
  Factors in Computing Systems}, pp. \bibinfo{pages}{1--7}.
\bibitem[{Sanchez-Vives and Slater(2005)}]{Sanchez2005presence}
\bibinfo{author}{Sanchez-Vives, M.V.}, \bibinfo{author}{Slater, M.},
  \bibinfo{year}{2005}.
\newblock \bibinfo{title}{From presence to consciousness through virtual
  reality}.
\newblock \bibinfo{journal}{Nature reviews neuroscience} \bibinfo{volume}{6},
  \bibinfo{pages}{332--339}.
\bibitem[{Schloss et~al.(2021)Schloss, Schoenlein, Tredinnick, Smith, Miller,
  Racey, Castro and Rokers}]{schloss2021uw}
\bibinfo{author}{Schloss, K.B.}, \bibinfo{author}{Schoenlein, M.A.},
  \bibinfo{author}{Tredinnick, R.}, \bibinfo{author}{Smith, S.},
  \bibinfo{author}{Miller, N.}, \bibinfo{author}{Racey, C.},
  \bibinfo{author}{Castro, C.}, \bibinfo{author}{Rokers, B.},
  \bibinfo{year}{2021}.
\newblock \bibinfo{title}{The uw virtual brain project: An immersive approach
  to teaching functional neuroanatomy.}
\newblock \bibinfo{journal}{Translational Issues in Psychological Science}
  \bibinfo{volume}{7}, \bibinfo{pages}{297}.
\bibitem[{Schubert et~al.(2001)Schubert, Friedmann and
  Regenbrecht}]{Schubert2001experience}
\bibinfo{author}{Schubert, T.}, \bibinfo{author}{Friedmann, F.},
  \bibinfo{author}{Regenbrecht, H.}, \bibinfo{year}{2001}.
\newblock \bibinfo{title}{The experience of presence: Factor analytic
  insights}.
\newblock \bibinfo{journal}{Presence: Teleoperators \& Virtual Environments}
  \bibinfo{volume}{10}, \bibinfo{pages}{266--281}.
\bibitem[{{SenseGlove}(2023)}]{SenseGloveNovaInfo}
\bibinfo{author}{{SenseGlove}}, \bibinfo{year}{2023}.
\newblock \bibinfo{title}{The new sense in vr for enterprise}.
\newblock
  \bibinfo{howpublished}{\url{https://www.senseglove.com/product/nova/}}.
\newblock \bibinfo{note}{[Online; accessed 8-March-2023]}.
\bibitem[{Staubli et~al.(2022)Staubli, Maloca, Kuemmerli, Kunz, Dirnberger,
  Allemann, Gehweiler, Soysal, Droeser, D{\"a}ster
  et~al.}]{staubli2022magnetic}
\bibinfo{author}{Staubli, S.M.}, \bibinfo{author}{Maloca, P.},
  \bibinfo{author}{Kuemmerli, C.}, \bibinfo{author}{Kunz, J.},
  \bibinfo{author}{Dirnberger, A.S.}, \bibinfo{author}{Allemann, A.},
  \bibinfo{author}{Gehweiler, J.}, \bibinfo{author}{Soysal, S.},
  \bibinfo{author}{Droeser, R.}, \bibinfo{author}{D{\"a}ster, S.}, et~al.,
  \bibinfo{year}{2022}.
\newblock \bibinfo{title}{Magnetic resonance cholangiopancreatography enhanced
  by virtual reality as a novel tool to improve the understanding of biliary
  anatomy and the teaching of surgical trainees}.
\newblock \bibinfo{journal}{Frontiers in Surgery} \bibinfo{volume}{9}.
\bibitem[{Steiniger et~al.(2022)Steiniger, Pfeffer, Gaffling and
  Lobachev}]{steiniger2022human}
\bibinfo{author}{Steiniger, B.S.}, \bibinfo{author}{Pfeffer, H.},
  \bibinfo{author}{Gaffling, S.}, \bibinfo{author}{Lobachev, O.},
  \bibinfo{year}{2022}.
\newblock \bibinfo{title}{The human splenic microcirculation is entirely open
  as shown by 3d models in virtual reality}.
\newblock \bibinfo{journal}{Scientific Reports} \bibinfo{volume}{12},
  \bibinfo{pages}{16487}.
\bibitem[{Syamlan et~al.(2022)Syamlan, Mampaey, Denis, Vander~Poorten,
  Tjahjowidodo et~al.}]{syamlan2022virtual}
\bibinfo{author}{Syamlan, A.}, \bibinfo{author}{Mampaey, T.},
  \bibinfo{author}{Denis, K.}, \bibinfo{author}{Vander~Poorten, E.},
  \bibinfo{author}{Tjahjowidodo, T.}, et~al., \bibinfo{year}{2022}.
\newblock \bibinfo{title}{A virtual spine construction algorithm for a
  patient-specific pedicle screw surgical simulators}, in:
  \bibinfo{booktitle}{2022 IEEE Symposium Series on Computational Intelligence
  (SSCI)}, \bibinfo{organization}{IEEE}. pp. \bibinfo{pages}{1493--1500}.
\bibitem[{Wang et~al.(2018)Wang, Song, Naqash, Zheng, Xu and
  Zhang}]{Wang2018TWH}
\bibinfo{author}{Wang, D.}, \bibinfo{author}{Song, M.},
  \bibinfo{author}{Naqash, A.}, \bibinfo{author}{Zheng, Y.},
  \bibinfo{author}{Xu, W.}, \bibinfo{author}{Zhang, Y.}, \bibinfo{year}{2018}.
\newblock \bibinfo{title}{Toward whole-hand kinesthetic feedback: A survey of
  force feedback gloves}.
\newblock \bibinfo{journal}{IEEE transactions on haptics} \bibinfo{volume}{12},
  \bibinfo{pages}{189--204}.
\bibitem[{Wang et~al.(2019)Wang, Yuan, Shiyi, Zhang, Weiliang and
  Jing}]{Wang2019HDF}
\bibinfo{author}{Wang, D.}, \bibinfo{author}{Yuan, G.}, \bibinfo{author}{Shiyi,
  L.}, \bibinfo{author}{Zhang, Y.}, \bibinfo{author}{Weiliang, X.},
  \bibinfo{author}{Jing, X.}, \bibinfo{year}{2019}.
\newblock \bibinfo{title}{Haptic display for virtual reality: progress and
  challenges}.
\newblock \bibinfo{journal}{Virtual Reality \& Intelligent Hardware}
  \bibinfo{volume}{1}, \bibinfo{pages}{136--162}.
\bibitem[{Xu et~al.(2021)Xu, Qiu, Jia, Dong, Yao, Xie, Guo, Yuan, Zhuang, Huang
  and Shi}]{medical_AI_VR}
\bibinfo{author}{Xu, X.}, \bibinfo{author}{Qiu, H.}, \bibinfo{author}{Jia, Q.},
  \bibinfo{author}{Dong, Y.}, \bibinfo{author}{Yao, Z.}, \bibinfo{author}{Xie,
  W.}, \bibinfo{author}{Guo, H.}, \bibinfo{author}{Yuan, H.},
  \bibinfo{author}{Zhuang, J.}, \bibinfo{author}{Huang, M.},
  \bibinfo{author}{Shi, Y.}, \bibinfo{year}{2021}.
\newblock \bibinfo{title}{Ai-chd: An ai-based framework for cost-effective
  surgical telementoring of congenital heart disease}.
\newblock \bibinfo{journal}{Commun. ACM} \bibinfo{volume}{64},
  \bibinfo{pages}{66–74}.
\newblock \URLprefix \url{https://doi.org/10.1145/3450409},
  \DOIprefix\doi{10.1145/3450409}.
\bibitem[{Zbontar et~al.(2018)Zbontar, Knoll, Sriram, Murrell, Huang, Muckley,
  Defazio, Stern, Johnson, Bruno et~al.}]{zbontar2018fastmri}
\bibinfo{author}{Zbontar, J.}, \bibinfo{author}{Knoll, F.},
  \bibinfo{author}{Sriram, A.}, \bibinfo{author}{Murrell, T.},
  \bibinfo{author}{Huang, Z.}, \bibinfo{author}{Muckley, M.J.},
  \bibinfo{author}{Defazio, A.}, \bibinfo{author}{Stern, R.},
  \bibinfo{author}{Johnson, P.}, \bibinfo{author}{Bruno, M.}, et~al.,
  \bibinfo{year}{2018}.
\newblock \bibinfo{title}{fastmri: An open dataset and benchmarks for
  accelerated mri}.
\newblock \bibinfo{journal}{arXiv preprint arXiv:1811.08839} .
\bibitem[{Zoller et~al.(2020)Zoller, Faludi, Gerig, Jost, Cattin and
  Rauter}]{zoller2020force}
\bibinfo{author}{Zoller, E.I.}, \bibinfo{author}{Faludi, B.},
  \bibinfo{author}{Gerig, N.}, \bibinfo{author}{Jost, G.F.},
  \bibinfo{author}{Cattin, P.C.}, \bibinfo{author}{Rauter, G.},
  \bibinfo{year}{2020}.
\newblock \bibinfo{title}{Force quantification and simulation of pedicle screw
  tract palpation using direct visuo-haptic volume rendering}.
\newblock \bibinfo{journal}{International journal of computer assisted
  radiology and surgery} \bibinfo{volume}{15}, \bibinfo{pages}{1797--1805}.
\bibitem[{Z{\"o}rnack et~al.(2021)Z{\"o}rnack, Weiss, Schummers, Eck and
  Navab}]{zornack2021evaluating}
\bibinfo{author}{Z{\"o}rnack, G.}, \bibinfo{author}{Weiss, J.},
  \bibinfo{author}{Schummers, G.}, \bibinfo{author}{Eck, U.},
  \bibinfo{author}{Navab, N.}, \bibinfo{year}{2021}.
\newblock \bibinfo{title}{Evaluating surface visualization methods in
  semi-transparent volume rendering in virtual reality}.
\newblock \bibinfo{journal}{Computer Methods in Biomechanics and Biomedical
  Engineering: Imaging \& Visualization} \bibinfo{volume}{9},
  \bibinfo{pages}{339--348}.

\end{thebibliography}

\end{document}